# Inside-Outside Duality for Planar Billiards – A Numerical Study

Barbara Dietz[1], Jean-Pierre Eckmann[2,3], Claude-Alain Pillet[2], Uzy Smilansky[4] and Iddo Ussishkin[4]

[1]University of Mexico (UNAM), Instituto de Física, Laboratorio de Cuernavaca, Apdo. Postal 139-B, 62191 Cuernavaca, Morelos, México
[2]Dépt de Physique Théorique, Université de Genève, CH-1211 Genève 4, Switzerland
[3]Section de Mathématiques, Université de Genève, CH-1211 Genève 4, Switzerland
[4]Dept. of Physics of Complex Systems, The Weizmann Institute of Science, 76100 Rehovot, Israel

**Abstract**

This paper reports the results of extensive numerical studies related to spectral properties of the Laplacian and the scattering matrix for planar domains (called billiards). There is a close connection between eigenvalues of the billiard Laplacian and the scattering phases, basically that every energy at which a scattering phase is $2\pi$ corresponds to an eigenenergy of the Laplacian. Interesting phenomena appear when the shape of the domain does not allow an extension of the eigenfunction to the exterior. In this paper these phenomena are studied and illustrated from several points of view.

We consider quantum billiards, i.e., the Laplacian in a bounded domain $\Omega$, with Dirichlet (zero) conditions on the boundary $\Gamma$. The billiard will be looked at from two different points of view, which define two seemingly independent problems. The interior problem is the more commonly studied aspect of the billiard dynamics, and the main objective in that case is to calculate the spectrum, i.e., the eigenvalues $E$ of the problem $-\Delta\psi = E\psi$, where $\psi$ vanishes on the boundary. In the exterior problem one considers scattering from the obstacle defined by the billiard boundary, with the same boundary conditions.

It was suggested [5] that there is a strong link between these two problems, which in a crude form states that an energy $E = k^2$ is an eigenenergy of the interior problem if and only if the on-shell scattering matrix $S(k)$ of the exterior problem has an eigenvalue 1. This statement is exact for the circular and elliptic billiards [3]. Using a truncated matrix, numerical calculation for the square [4] gives excellent agreement. In the semiclassical limit, it is justified by observing that the semiclassical spectral density it predicts [5] coincides with the Gutzwiller trace formula [7].

As we shall see below, the conjecture implies that at an eigenenergy $E_n$ the obstacle is "transparent" for a well-chosen wave function. In the interior of the billiard it equals the eigenfunction of the interior problem, and in the exterior of the billiard it is the wave function corresponding to the eigenvalue 1 of the scattering matrix. The conjecture therefore leads to the result that the eigenfunction of the interior problem can be continued to a single valued bounded function in the plane. Billiards whose eigenfunctions may not be continued to the whole plane, due to branch points, are easily constructed [6]. The "cake" billiard discussed in that paper is one of the examples, and other examples without corners are also given.

These examples show that the conjecture cannot hold in the form stated above, but a rephrasing of the basic idea leads to the following relation between the inside problem and the scattering problem. We consider domains which are simply connected[1], with a boundary which is piecewise

[1]In fact, the domain can also have several pieces, provided that the complement of $\Omega$ is connected.

$\mathcal{C}^2$ (i.e., twice differentiable) and with the corner angles bounded away from 0 and $2\pi$. We also introduce the eigenvalues of the (unitary) scattering matrix $S(k)$ for the exterior problem (with Dirichlet boundary conditions), restricted to the energy shell $E = k^2$: They are denoted by $e^{-i\theta_j(k)}$, with $\theta_j(k) \in [0, 2\pi)$, and the $\theta_j(k)$ are called the scattering phases[2]. We finally denote by $\Delta_\Omega$ the Laplacian in $\Omega$ with Dirichlet boundary conditions on $\Gamma$. The following results were recently proved in [6]. The general arrangement of the eigenvalues is described by

**Lemma 1** *Let $k > 0$. The on-shell scattering matrix $S(k)$ for the exterior domain, with Dirichlet boundary conditions on $\Gamma$, is unitary. Its eigenvalues accumulate only at 1. They accumulate from the lower half-plane only.*

The following theorem provides the precise form of what we call the inside-outside duality.

**Theorem 1** *The following two statements are equivalent:*

1. $-\Delta_\Omega$ *has an $m$-fold degenerate eigenvalue $k_0^2$.*
2. *As $k \uparrow k_0$, exactly $m$ scattering phases of the scattering matrix $S(k)$ converge to $2\pi$ from below.*

The case of an eigenvalue 1 for the $S$-matrix is more special and is closely related to the extension of the inside eigenfunction to the whole plane:

**Theorem 2** *Assume that for $k_0 > 0$ the scattering matrix $S(k_0)$ has an eigenvalue 1 of multiplicity $m$. Then $-\Delta_\Omega$ has an eigenvalue $k_0^2$ of multiplicity at least $m$. The corresponding eigenfunctions can be extended to bounded solutions of the Helmholtz equation in* $\mathbf{R}^2$.

The main element in this formulation of the inside-outside duality is the realization that although one of the eigenphases approaches $2\pi$ (from below) as $k \uparrow k_0$, 1 is not necessarily an eigenvalue of $S(k_0)$. The reason why this can happen lies in the accumulation of eigenvalues of $S(k)$ at 1 from the lower half-plane, which can cause intricate interactions between many different scattering phases.

One of the aims of the present paper is to describe with a few numerical examples some consequences of the interaction mechanism which is at work when $k$ approaches $k_0$ from below. They will show in what sense the obstacle is “transparent”, and explain why semiclassical approximations for the interior problem, which use the inside-outside duality, produce useful results. The detailed study of these examples enables also a better understanding of the advantages and the accuracy obtained in numerical work which is based on the duality of the exterior and the interior problems. Finally, we shall argue that despite the interactions between the scattering phases, the nature of the evolution of the spectrum of $S(k)$ can be partially explained on physical grounds, adding thereby to the information provided by the rigorous results.

We divide the numerical results into two parts. The crude form of the inside-outside duality suggests that the solution of the billiard problem may be expanded in plane waves (at the same energy). In Section 2, we examine what happens if such an expansion is made (despite our conclusion that it cannot exist, generally). The results of these calculations are closely related to those of Berry [2]. In Section 3 we examine the vicinity of the eigenenergy. The behavior of the scattering phases which are close to 0 is examined in this region, as well as the properties of the corresponding scattering solutions. Finally, we discuss how it can happen that while the original conjecture is not precise, yet semiclassical approximations for the interior problem may be derived using it, and yield extremely precise results.

[2]In the paper [6], the eigenvalues of $S(k)$ are defined as $e^{-2i\theta_j}$.

# 1 The Examples and Numerical Considerations

In this section, we discuss the shapes of the domains considered in this paper, and the nature of the numerical schemes which will be used.

The crude form of inside-outside duality is correct for the *circular* billiard. For a circle with a radius $a$ the interior eigenstates (in polar coordinates) are

$$\psi_{\ell,n}(r,\varphi) \;=\; J_\ell(k_{\ell,n} r) e^{i\ell\varphi} \;, \tag{1}$$

where $J_\ell(x)$ denotes the Bessel function of order $\ell$, and the numbers $k_{\ell,n}a$ are its zeros. The scattering matrix $S(k)$ is diagonal in the angular momentum representation, $S_{\ell,\ell'}(k) = S_\ell(k)\delta_{\ell,\ell'}$, with

$$S_\ell(k) \;=\; e^{-i\theta_\ell(k)} \;=\; -\frac{H^-_\ell(ka)}{H^+_\ell(ka)} \;, \tag{2}$$

where $H^\pm_\ell(x) = J_\ell(x) \pm iY_\ell(x)$ denotes the Hankel functions. An eigenvalue assumes the value 1 when $J_\ell(ka) = 0$, which is exactly the quantization criterion for the interior problem. At these energies, the interior wave functions can be extended outside and written as a superposition of plane waves for which the obstacle is transparent. The cylindrical wave is expressed in terms of plane waves by

$$J_\ell(kr)e^{i\ell\varphi} \;=\; \frac{i^{-\ell}}{2\pi}\int_0^{2\pi} e^{i\mathbf{k}\cdot\mathbf{r}} e^{i\ell\phi} d\phi \;, \tag{3}$$

where, in polar coordinates $\mathbf{k}\cdot\mathbf{r} = kr\cos(\varphi-\phi)$. Expansions in terms of the cylindrical functions $J_\ell(kr)e^{i\ell\varphi}$ will therefore also be called expansions in plane waves.

The *square* billiard is another example where the inside eigenfunctions may be continued to bounded solutions of the Helmholtz equation in the plane. For a square with corners at $(\pm a, \pm a)$, the eigenfunctions are given by

$$\psi_{m,n}(x,y) \;=\; \sin\left(\frac{\pi m}{2a}(x+a)\right)\sin\left(\frac{\pi n}{2a}(y+a)\right) \;, \tag{4}$$

with energy $E = (m^2+n^2)\pi^2/(4a^2)$. While the inside problem is separable, the scattering problem is not [4]. The scattering matrix does not have an analytic expression. It is nevertheless clear that it does have a generalized eigenvalue 1 at an eigenenergy of the interior problem, and the corresponding eigenvector may be found using the continuation of the interior eigenfunction. (The eigenfunction is only generalized, because it is a sum of 4 $\delta$-functions on the energy shell and hence not normalizable in $L^2$.) Berry [2] has exhibited domains where the exterior functions grow at infinity, and therefore their restriction to the energy shell can again not be defined properly.

The situation ceases to be so simple for the *cake* billiard, defined as a sector of the circle:

$$\Omega = \left\{(r,\varphi)\;:\;0<r<a,\;|\varphi|<\frac{\alpha}{2}\right\}\;. \tag{5}$$

Its eigenfunctions are given by

$$\psi_{\ell,n}(r,\varphi) \;=\; \sin\left(\frac{\pi\ell}{\alpha}\left(\varphi+\frac{\alpha}{2}\right)\right) J_{\frac{\pi\ell}{\alpha}}(k_{\ell,n}r)\;, \tag{6}$$

where the $k_{\ell,n}$ satisfy the quantization condition

$$J_{\frac{\pi\ell}{\alpha}}(k_{\ell,n}a) = 0 . \tag{7}$$

If $\pi\ell/\alpha$ is not an integer, the solutions $\psi_{\ell,n}$ have a branch point at the origin and are not single valued functions in the plane. Thus, the exterior function is ill defined, and by Theorem 2, the number 1 is not an eigenvalue of $S(k_{\ell,n})$ and therefore the crude form of the inside-outside duality cannot hold in this case.

The numerical calculations were done for the cake (with $\alpha = 2\pi/3$, $a = 1$), the square (with $a = 1$) and the stadium (with a ratio of 1.8 between the length of the straight segment and the diameter of the circle). Of course, we are forced to truncate the calculation to a finite dimension. Our numerical approximations for the scattering matrix take advantage of the compactness of the obstacle. We truncate the scattering matrix to a size $\Lambda = \Lambda_{sc} + \Lambda_e$ in the angular momentum basis, where $\Lambda_{sc}$ is the number of semiclassically relevant eigenvalues. It is estimated semiclassically (for convex billiards) as [3]

$$\Lambda_{sc} = \left[\frac{k|\Gamma|}{\pi}\right] , \tag{8}$$

where $\Gamma$ is the boundary of the billiard, and $[\cdots]$ denotes the integer part. The number $\Lambda_e$ of waves with higher angular momentum (evanescent waves) we have chosen in our calculations depends on the question we ask, and will be explained in each case.

## 2 The Plane Wave Expansion at an Eigenenergy

The inside-outside duality is based on the idea that at an eigenenergy of the interior problem the corresponding eigenfunction may be expanded in plane waves (at the same energy) to the whole plane. If this expansion exists, it will then be equal to the eigenfunction inside the billiard, while outside the billiard it will be equal to a scattering solution corresponding to the eigenvalue 1 of the $S$-matrix. The general problem of the expansion of eigenfunctions in terms of plane waves is abundantly treated in the literature, see, e.g., [1, 8].

The numerical calculations of this section examine what happens if such an expansion is made in a case where the outside function does exist (the square) and for the more interesting case where it does not (the cake). For a similar study, see also the results of Berry [2].

The calculations in this section are done for the lowest energy level. The angular momentum representation is used, and working with the lowest energy level enables us to use a minimal number of components in the various expansions, while still including a sufficient number of evanescent modes. The symmetries of each billiard were also used to reduce calculations.

We expand the wave function in the form

$$\psi(r,\varphi) = \sum_{\ell=-\infty}^{\infty} a_\ell J_\ell(k_0 r) e^{i\ell\varphi} , \tag{9}$$

where $k_0^2$ is the eigenenergy.

For the ground state of the *square* one has

$$\psi_0 = \cos\left(\frac{k_0 x}{\sqrt{2}}\right)\cos\left(\frac{k_0 y}{\sqrt{2}}\right) = \sum_{\ell=-\infty}^{\infty} (-1)^\ell J_{4\ell}(k_0 r) e^{4i\ell\varphi} , \tag{10}$$

where $k_0^2 = \pi^2/2a^2$, so that the coefficients in (9) are given by

$$a_\ell = \begin{cases} (-1)^\ell, & \ell = 4n \\ 0, & \text{otherwise} \end{cases} . \tag{11}$$

Note that for the ground state of the square, one finds $\Lambda_{\text{sc}} = [4\sqrt{2}] = 5$.

For the *cake* an expansion is not known analytically. To get a numerical approximation, the sum (9) is truncated at $\ell = \pm L$, and the coefficients $a_\ell$ are determined by demanding the expansion and the ground state

$$\psi_0(r, \varphi) = \cos\left(\frac{\pi\varphi}{\alpha}\right) J_{\frac{\pi}{\alpha}}(k_0 r) \tag{12}$$

to have the same normal derivative on the boundary. We use the following method: Both normal derivatives are expanded in terms of a complete set on the boundary, and the coefficients of this expansion are compared. The symmetry around the $x$-axis is used. Note that for the ground state of the cake, one finds $\Lambda_{\text{sc}} = [(2/\pi + \pi/3)k_0] = 5$ and $\Lambda = 2L + 1$.

Contour plots of the resulting functions are given for the square in Fig. 1, and for the cake in Fig. 2. The plots for the square where obtained using exactly the same procedure as for the cake, with the resulting coefficients being close to the exact ones (11).

For the square, as $\Lambda$ is increased, the expansion approaches the correct function in the plane in a growing region around the origin. In particular, inside the square it approaches the eigenfunction itself. These results are not surprising in view of the expansion (10), but are shown for comparison with the cake.

For the cake, the behavior of the expansion is very different for the inside and the outside. This is already visible in Fig. 2. For the *inside* of the cake, the various figures will reveal convergence, while for the outside we see higher and higher oscillations. These findings are illustrated in more detail in Fig. 3. In (a), we show the nodal lines of the expansion described above for increasing $\Lambda$. We see that they approach the boundary of the cake. The nature of the convergence to the upper corner might suggest general problems for domains with corners. However, the problems with the cake are of a different nature, due to the branch point in the outside continuation of the eigenfunction. To make this statement more precise, we consider the cuts as indicated in (b). In the plot (c) it is manifest that in the inside (the segment D-E) the approximations converge to the exact wave function Eq.(12). For the *outside*, we see in Fig. 2 an increasing number of oscillations, as $\Lambda$ is increased. This becomes quantitatively clearer in Fig. 3. In (c), we find that the function diverges faster in the radial direction (once outside the cake) as $\Lambda$ grows. In the logarithmic representation (d) of the absolute value of the functions along the cake, one sees the simultaneous growth of the amplitude and an increase of the oscillations. All these features indicate that the approximation converges weakly to zero as $\Lambda \to \infty$.

Another aspect of the irregular behavior outside the cake is illustrated in Fig. 4 by the size of the coefficients in the plane wave expansion, which are seen to behave like an exponentially growing sequence as $\Lambda$ increases. Thus, the plane wave expansion clearly does not converge in the exterior.

## 3 The Scattering Problem Near an Eigenenergy

In the previous section, we studied the "difficulties" which occur if one tries to construct the outside wave function when it does not exist. In this section, we look at the same problem from another angle, namely, we analyze what happens to the scattering phases and wave functions as $k$ crosses a value $k_0$ corresponding to an interior eigenvalue $k_0^2$.

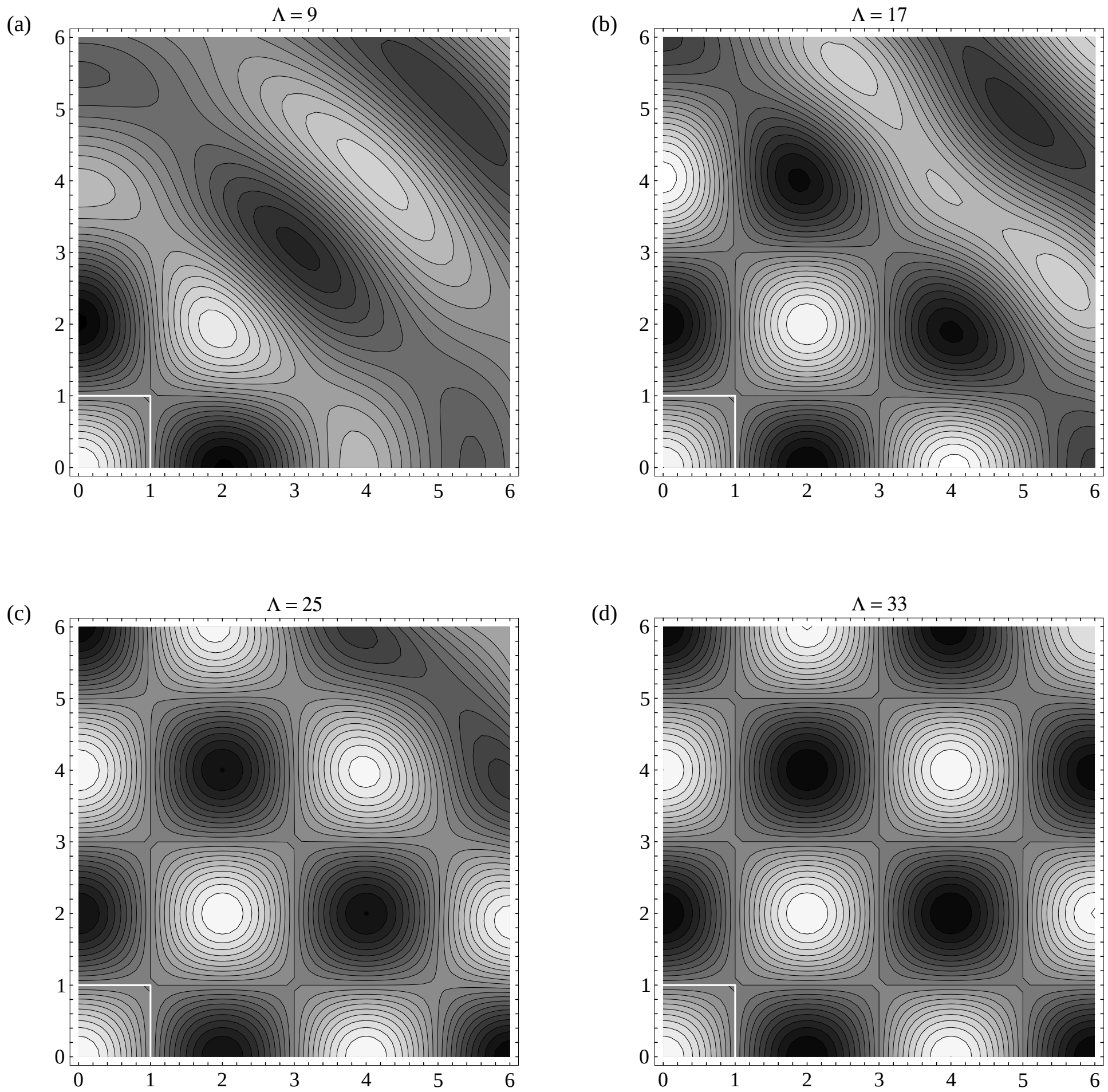

Fig. 1: The eigenfunction of the square as a finite expansion, for $\Lambda = 9, 17, 25, 33$. The shades of gray interpolate between the extremal values of $\pm 1$ of the approximation to the function Eq.(10). The white line is the boundary of a quarter of the square, and the figures extend symmetrically around both axes.

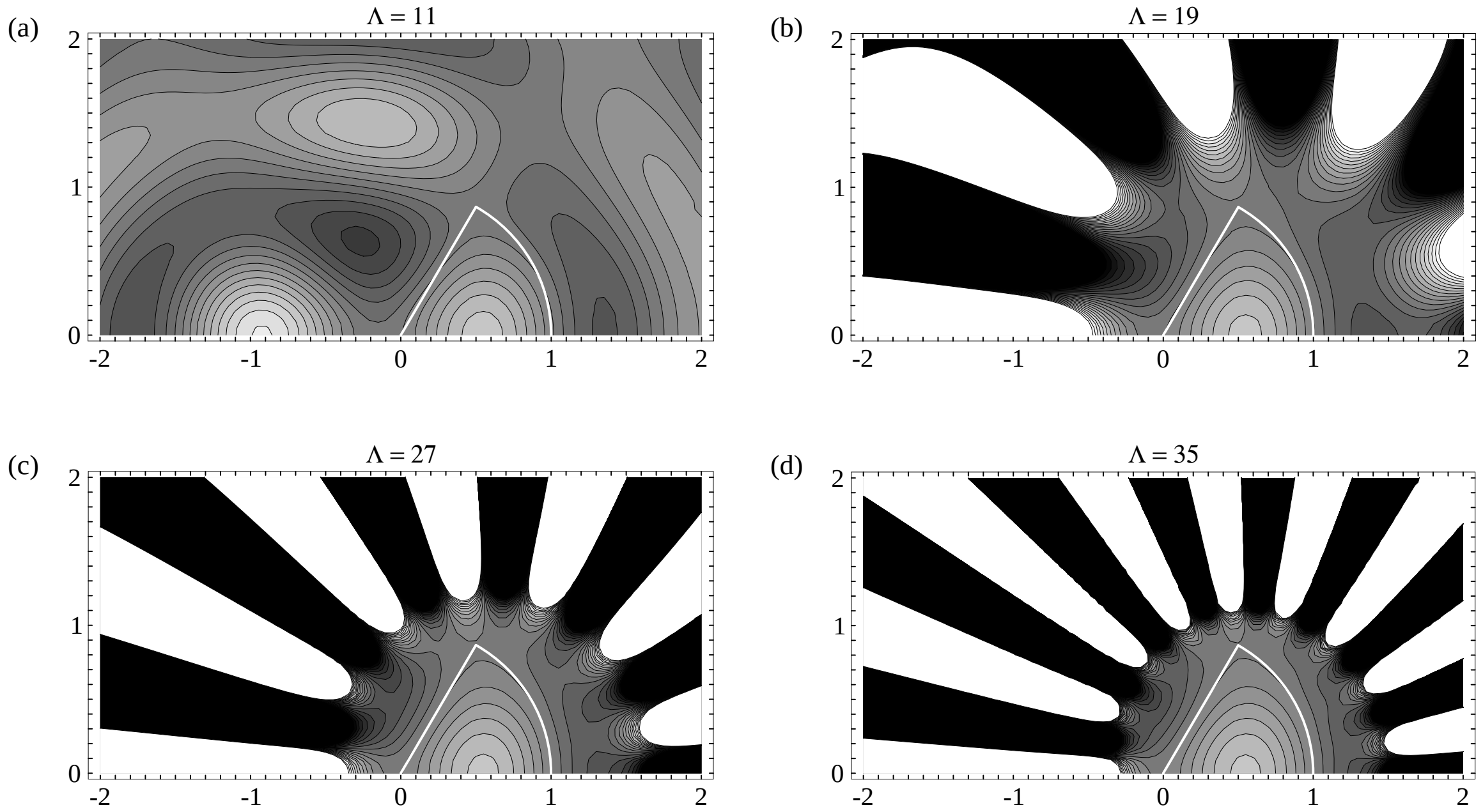

Fig. 2: The eigenfunction of the cake as a finite expansion, for $\Lambda = 11, 19, 27, 35$. The shades of gray interpolate between the values of $\pm 1$ of the expansion. Larger values are black or white. The white line is the boundary of the cake of which only the upper half is shown. The figures extend symmetrically below the $x$-axis.

As in the previous section, a truncation to size $\Lambda$ of the $S$-matrix is used. The scattering matrix of the cake is calculated in the following manner: To reduce the amount of computation, all calculations have been done in the subspace of functions which are symmetric around the $x$-axis. This means that we only look at the symmetric block of the scattering matrix. The wave function for a given scattering solution is expressed for $r > 1$ as

$$\psi_\ell^{(\text{far})}(r,\varphi) \;=\; H_\ell^-(kr)\cos(\ell\varphi) + \sum_{\ell'=0}^{\infty} S_{\ell,\ell'} H_{\ell'}^+(kr)\cos(\ell'\varphi)\;, \tag{13}$$

and for $r < 1$ (and $\varphi \in [\alpha/2, 2\pi - \alpha/2]$) as

$$\psi_\ell^{(\text{near})}(r,\varphi) \;=\; \sum_{n=0}^{\infty} a_{\ell,n} \cos\left(\left(n+\tfrac{1}{2}\right)\pi\,\tfrac{\pi-\varphi}{\pi-\frac{\alpha}{2}}\right) J_{\frac{(2n+1)\pi}{2\pi-\alpha}}(kr)\;. \tag{14}$$

Writing the functions in this way ensures symmetry around the $x$-axis, and $\psi = 0$ on the straight segments of the billiard. Note that the regions have a common arc, on which these functions will be matched. In the numerical calculation the sums are truncated to $\ell' = 0, \ldots, L'$ in Eq.(13) and to $n = 0, \ldots, N$ in Eq.(14). The matching conditions which determine the coefficients are found by expanding $\psi_\ell^{(\text{far})}$, $\psi_\ell^{(\text{near})}$, and their normal derivatives, in a complete set of functions on the billiard's arc ($r = 1, \varphi \in [-\alpha/2, \alpha/2]$), and on the common arc ($r = 1, \varphi \in [\alpha/2, 2\pi - \alpha/2]$). We determine $L' + N + 2$ components of the expansion by requiring

$$\begin{aligned} \psi_\ell^{(\text{far})} &= 0 && \text{on the billiard's arc,} \\ \psi_\ell^{(\text{far})} &= \psi_\ell^{(\text{near})} && \text{on the common arc,} \\ \partial_{\mathbf{n}} \psi_\ell^{(\text{far})} &= \partial_{\mathbf{n}} \psi_\ell^{(\text{near})} && \text{on the common arc.} \end{aligned} \tag{15}$$

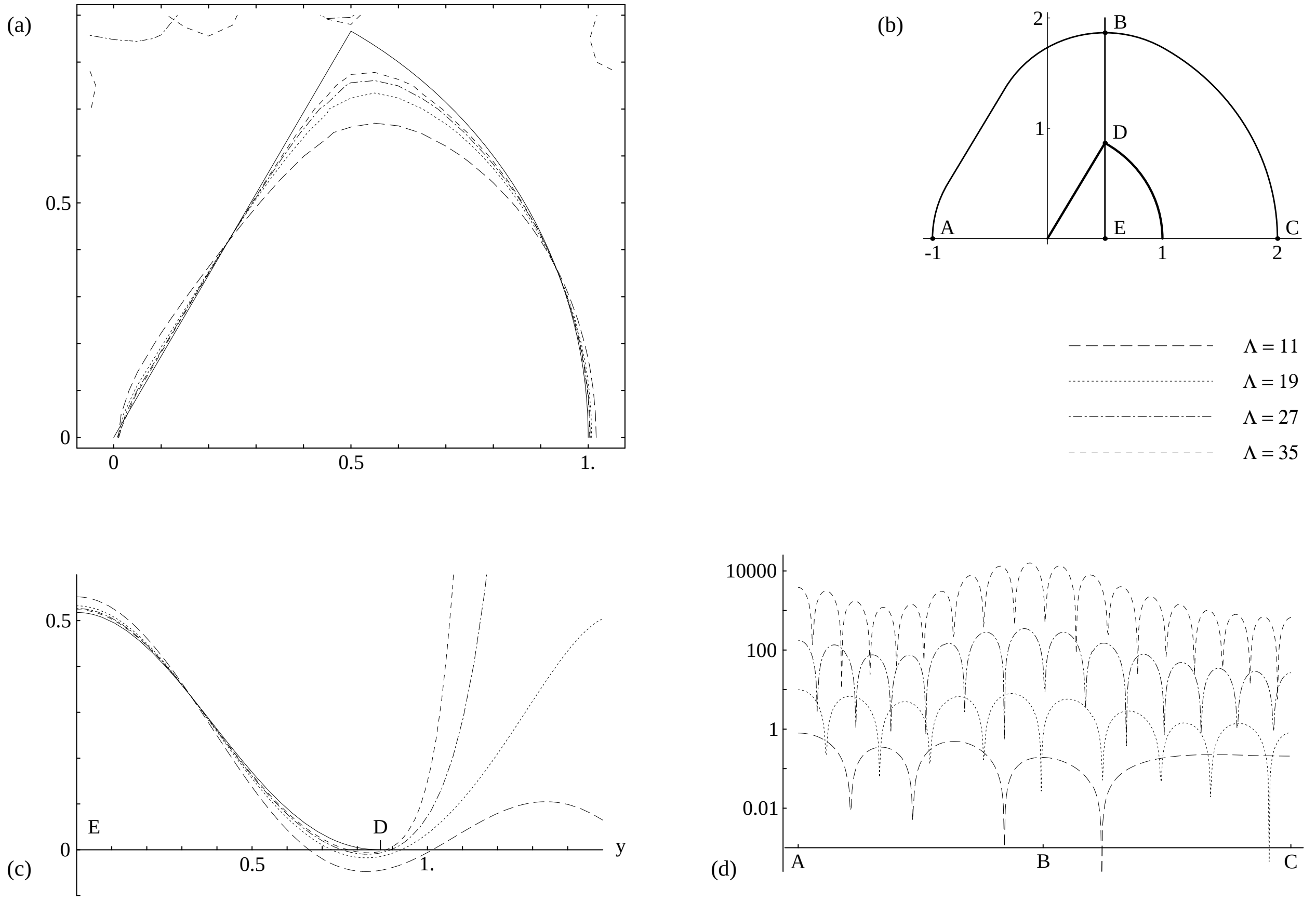

Fig. 3: (a) The nodal line of the eigenfunction for the cake for a finite expansion, with $\Lambda = 11, 19, 27, 35$. (b) Position of the cuts shown in (c) and (d). (c) The approximate wave function along the cut B-D-E, depending on $\Lambda$. Note the convergence to the exact function (shown as a solid line) on the segment D-E. (d) Logarithmic representation of the absolute value of the approximation along the segment A-B-C.

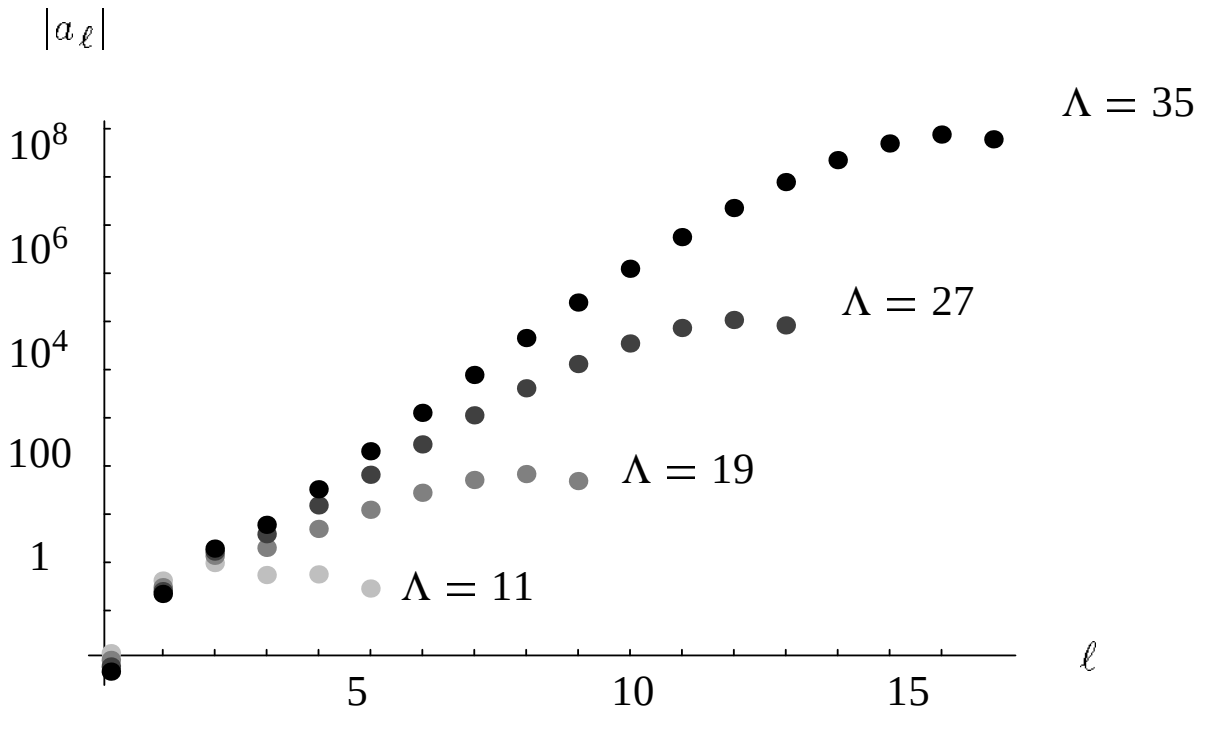

Fig. 4: Coefficients $a_\ell$ of the plane wave expansion, Eq. (9), for $\Lambda = 11, 19, 27, 35$.

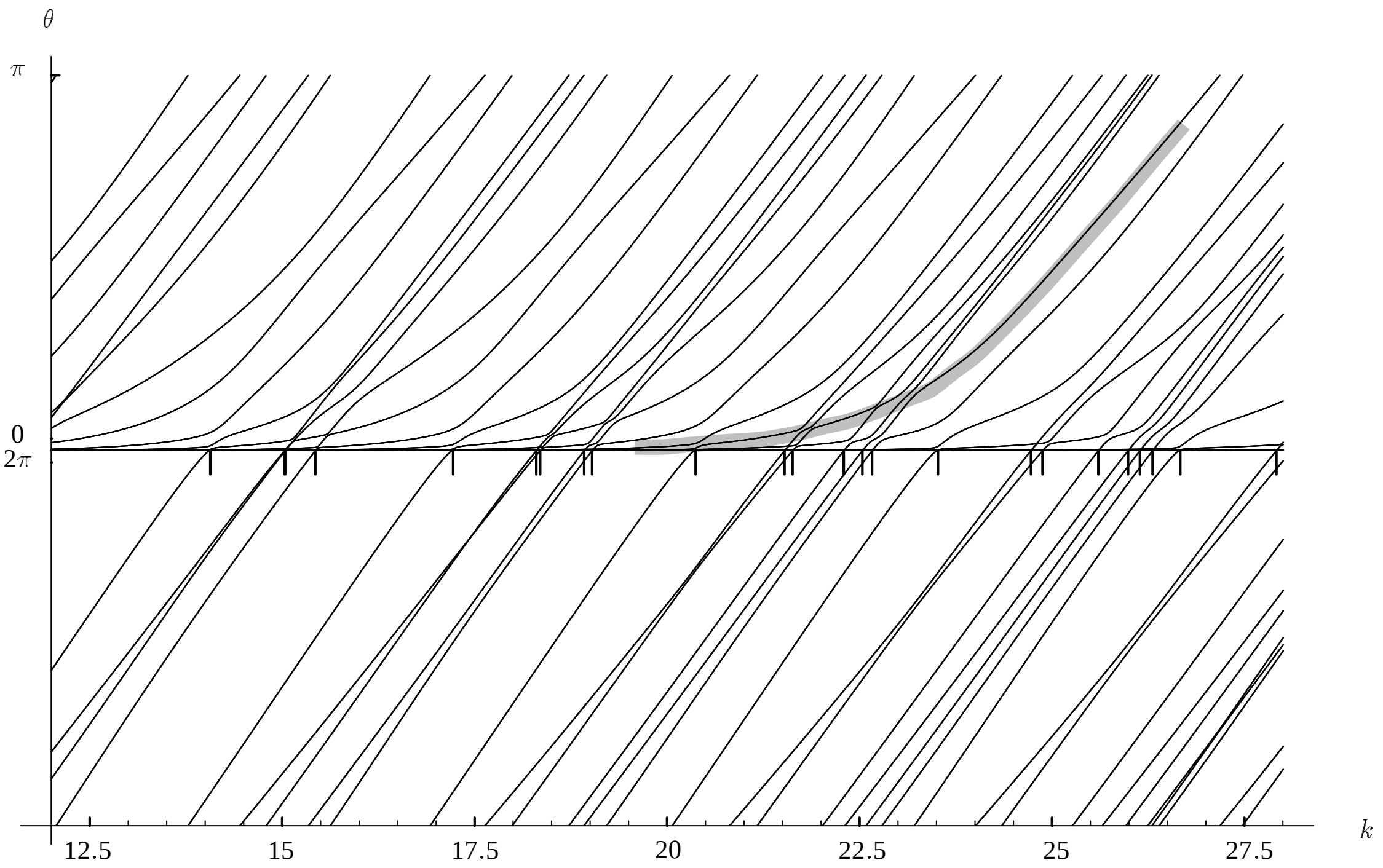

Fig. 5: The scattering phases as a function of $k$ for the case of the cake. The calculation was done with $L' = 25$ and $N = 16$. Note that whenever a scattering phase reaches the central line from below at $k$, then $k^2$ corresponds to an eigenvalue of the inside problem, as indicated by the vertical lines. The gray overlay shows a new semiclassically relevant channel.

This procedure gives a truncated scattering matrix of size $L' + 1$. It is not exactly unitary, but with increasing $L'$ the deviation from unitarity decreases and the eigenvalues converge.

The method which was described above for the calculation of the $S$-matrix for the cake boundary is optimal for this particular problem, because it takes advantage of the special symmetry and shape of the cake. In general, it is advantageous to use another method, which expresses the $S$-matrix in terms of boundary integrals. This method is known as the "null field method" in the literature (see e.g., Martin [10] and refs. cited therein). For star-like boundaries, one can use a simpler version, which is derived and explained in [3]. There, it was used to study the ellipse billiards, yielding the correct scattering phases.

Our first illustration, Fig. 5, shows the scattering phases (always on the symmetric subspace) for $k \in [12, 28]$, where the phase $\theta = 0 = 2\pi$ is placed at the center of the figure. The inside-outside duality (Theorem 1) means that whenever an eigenphase reaches the central line from below at $k$, then $k^2$ corresponds to an eigenvalue of the inside problem. The eigenvalues are shown with vertical bars, and the coincidence is manifest. The first such picture was presented in the paper [4] for the case of the square, and it suggested the inside-outside theorem stated above. The gray band corresponds to a new semiclassically relevant channel [12]. The density of opening channels in the figure is predicted by the semiclassical formula Eq.(8).

A closer look at Fig. 5 reveals that the scattering phases seem to cross the horizontal line $\theta = 2\pi$, despite the rigorous result that the eigenvalue ceases to exist for the cake when $\theta = 2\pi$. The next calculation will analyze the nature of this crossing for the lowest eigenvalue of the cake. In order to get precise results, these calculations were done with $L' = 22$, $N = 14$ and carrying 61 digits of numerical precision to avoid roundoff errors. We have already seen that $\Lambda_{\text{sc}} = 5$. Since $L' = 22$, this means that a large number of evanescent modes are carried along. This avoids truncation errors and numerical tests show that the scattering phases can be trusted at least up to an absolute precision of $10^{-6}$, and also to a relative precision which deteriorates to a maximum of about 25 percent when the eigenphase is $10^{-15}$.

The results of this high-precision calculation near $k = k_0$ are summarized in Figs. 6 and 7. The truncation allows one to compute a total of 23 phases. In Fig. 6(a), 5 phases are too large to be seen, while 16 phases are too small to be distinguished from 0, and the behavior of only 2 phases may be studied at the graph's scale. Consider the scattering phase which is negative for $k < k_0$. Looking at this graph only, one might think the phase does become 0, with a small error in $k$ due to the truncation of the scattering matrix. It also seems to continue to a positive phase for $k > k_0$.

However, closer inspection reveals a much more subtle "avoided crossing" picture near $k_0$. We can already interpret graph (a) as such an avoided crossing, with the lower phase repelling the one above the $k$-axis. As seen in graphs (b)–(f), this picture repeats in many successive enlargements (in $\theta$), in such a way that one cannot really tell one graph from another. In particular, what looked like a crossing in graph (a) is in reality an avoided crossing which becomes visible in graph (b), and what seems to be a crossing at some enlargement turns out to be an avoided one when one magnifies further. In the Fig. 7, 12 phases on the side $\theta > 0$ are shown on a logarithmic scale, and the sequence of a large number of avoided crossings reappears nicely.

Thus, we realize that what looks like a crossing in Fig. 5 is in reality an infinite sequence of avoided crossings. For the understanding of these crossings, it is important to note that the eigenphase coming from below *repels distant positive eigenphases earlier than those which are closer to* $\theta = 0$. In Fig. 6(a), the interaction is between the two distant phases, although there is an infinite number of other phases between them, which are slowly revealed under the enlargements (b)–(f).

To understand the interaction between the various eigenvalues, one can analyze the simplified problem of a domain which is a small perturbation of the circle. For the circle, the $S$-matrix is diagonal in the angular momentum basis, and for $\ell \gg ka$, where $a$ is the radius of the circle, the scattering phases are well approximated by const. $\ell(ka/2)^{2\ell}/\ell!^2$. Since the problem has rotational symmetry there is no interaction between the phases, and they cross each other as $k$ is varied. Introducing a deformation of the circle, the $S$-matrix is perturbed, and one can estimate the size of the off-diagonal matrix elements. In the angular momentum basis one finds

$$S_{\ell,\ell'} = \text{const.} \left(\frac{ka}{2}\right)^{\ell+\ell'} \frac{1}{(\ell-1)!\,(\ell'-1)!}\, g_{\ell-\ell'} \,, \tag{16}$$

if both $\ell \gg ka$ and $\ell' \gg ka$. If $\ell \gg ka$ but $\ell'$ is in the semiclassical region, then one gets another

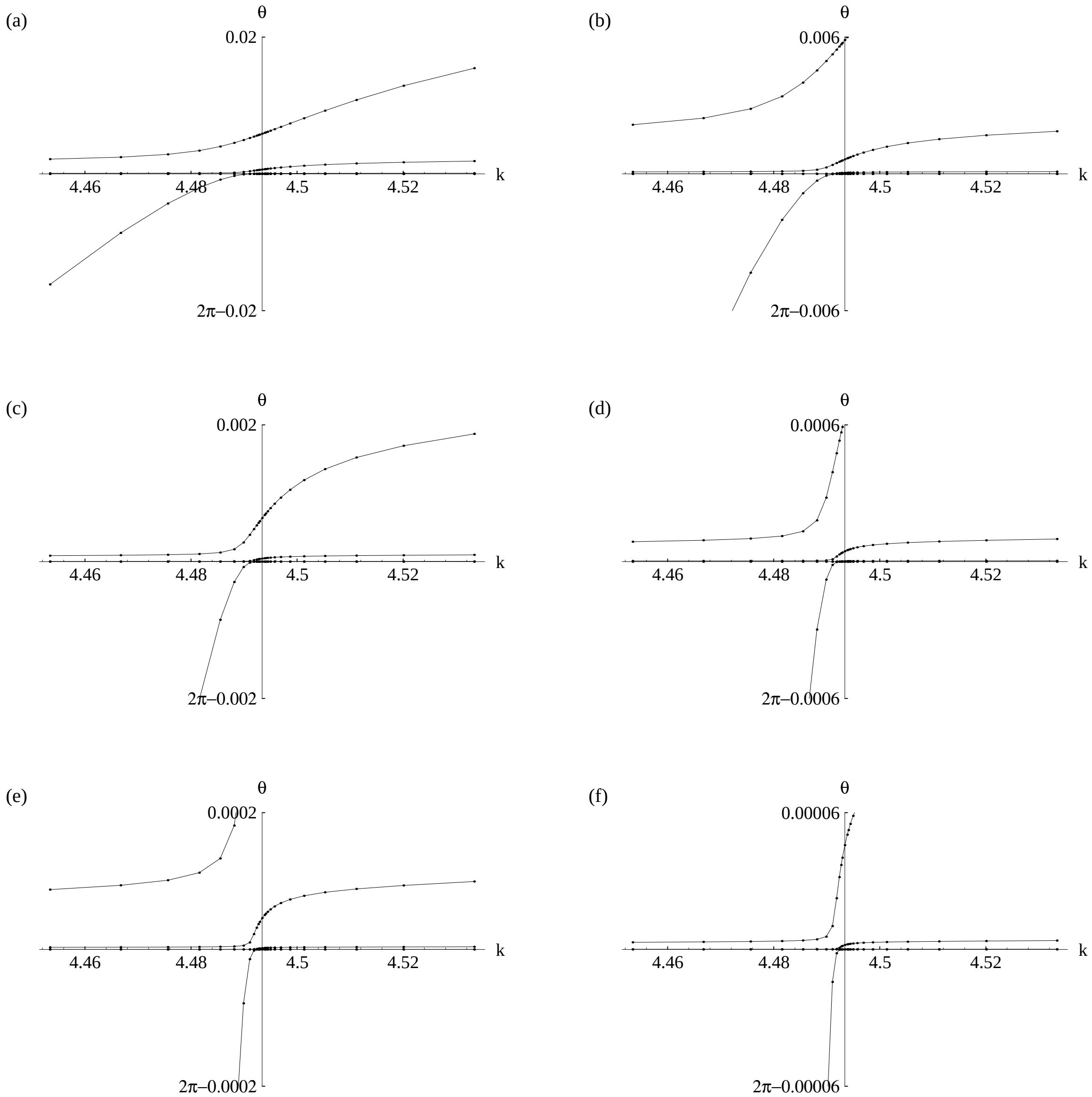

Fig. 6: Scattering phases of the cake drawn at different scales. The phase axis is placed at $k_0$, which corresponds to the lowest eigenvalue of the cake.

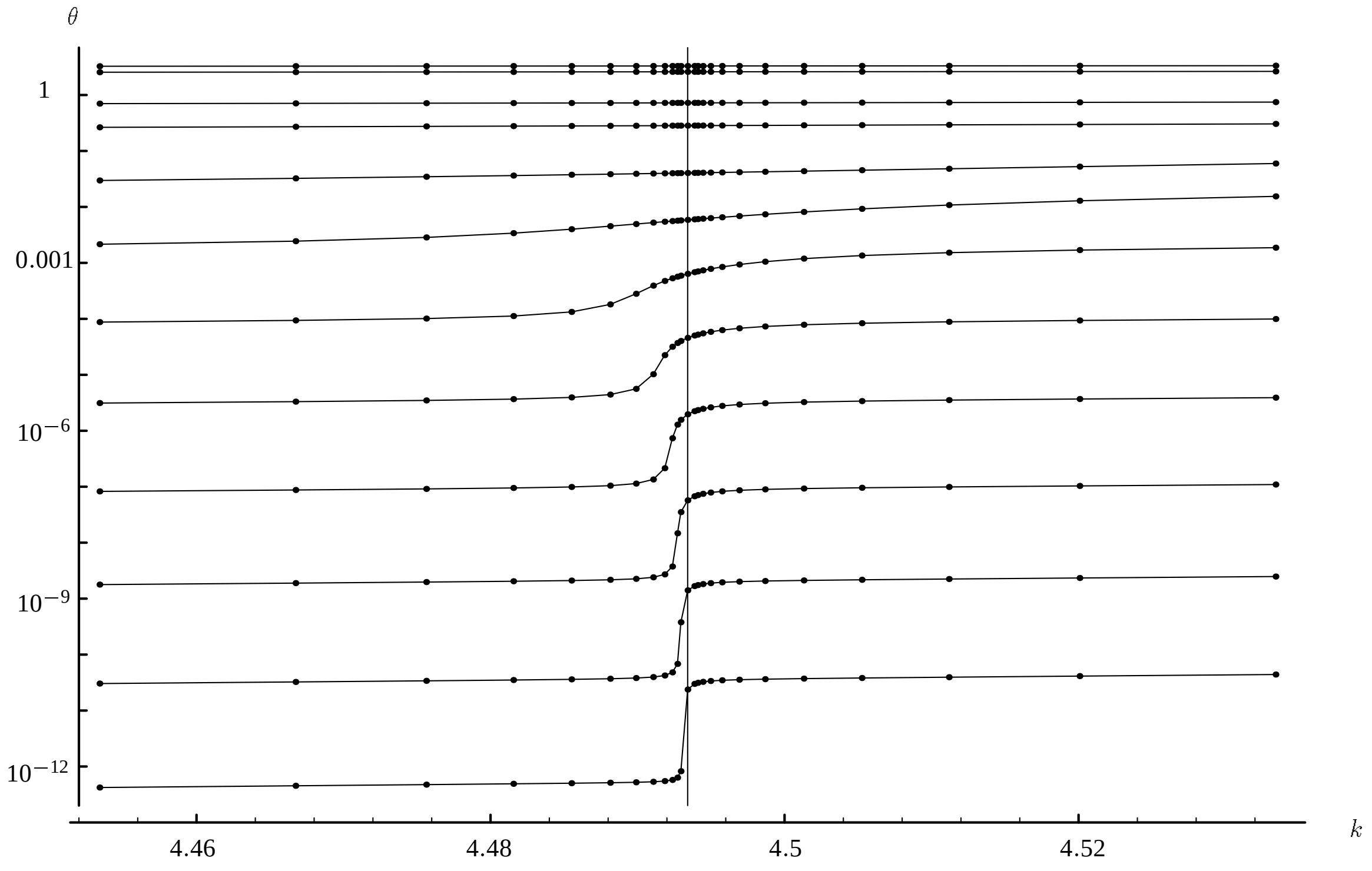

Fig. 7: The same data as in Fig. 6, but only for $\theta > 0$, shown on a logarithmic scale.

formula, namely

$$S_{\ell,\ell'} \;=\; \text{const.} \left(\frac{ka}{2}\right)^{\ell-\frac{1}{2}} \frac{1}{(\ell-1)!}\, g_{\ell-\ell'} \;, \tag{17}$$

where the correcting factor $g_{\ell-\ell'}$ in Eq.(16) and (17) is the Fourier component of the deformation. It decays exponentially with $|\ell - \ell'|$ for an analytic perturbation of the circle.

In analogy with the formulas for the avoided crossing of two eigenvalues, we conclude that two phases are perturbed substantially if $|S_{\ell,\ell'}|^2/|(\theta_\ell - \theta_{\ell'}) \bmod 2\pi|$ is large. Using this estimate, it is seen that as one of the semiclassical phases approaches $2\pi$, it will interact first with the evanescent phase with lowest $\ell$, as can be seen in Fig. 6. Explaining further details of the successive crossings is beyond this simple analysis.

In a two-level avoided crossing, the two eigenvalues "exchange" their identities. In the multiple avoided crossings for the cake, the angular momentum is permuted among the various levels as they interact. This can be seen by looking at the eigenfunction corresponding to that scattering phase which reaches $2\pi$ as $k \uparrow k_0$. This eigenfunction must cease to exist for the cake at $k = k_0$. In Fig. 8 contour plots of the absolute value of the scattering solution are given for $k$ approaching $k_0$. These solutions are normalized to produce a $k$-independent incoming flux. Using Eqs.(13)–(14), they are given by $\psi = \sum_\ell v_\ell \psi_\ell$, where $v$ is the (normalized) left eigenvector of $S$ corresponding to the scattering phase which approaches $2\pi$, with the normalization $\sum_\ell |v_\ell|^2 = 1$. We note two observations.

First, it is clear from the plots that as $k$ approaches $k_0$ the larger values of $\ell$ become dominant, indicating the "exchange" of angular momentum. The last plot resembles Fig. 2(d): both plots are dominated by high angular momenta. This is also visible in Fig. 9 where we show the average

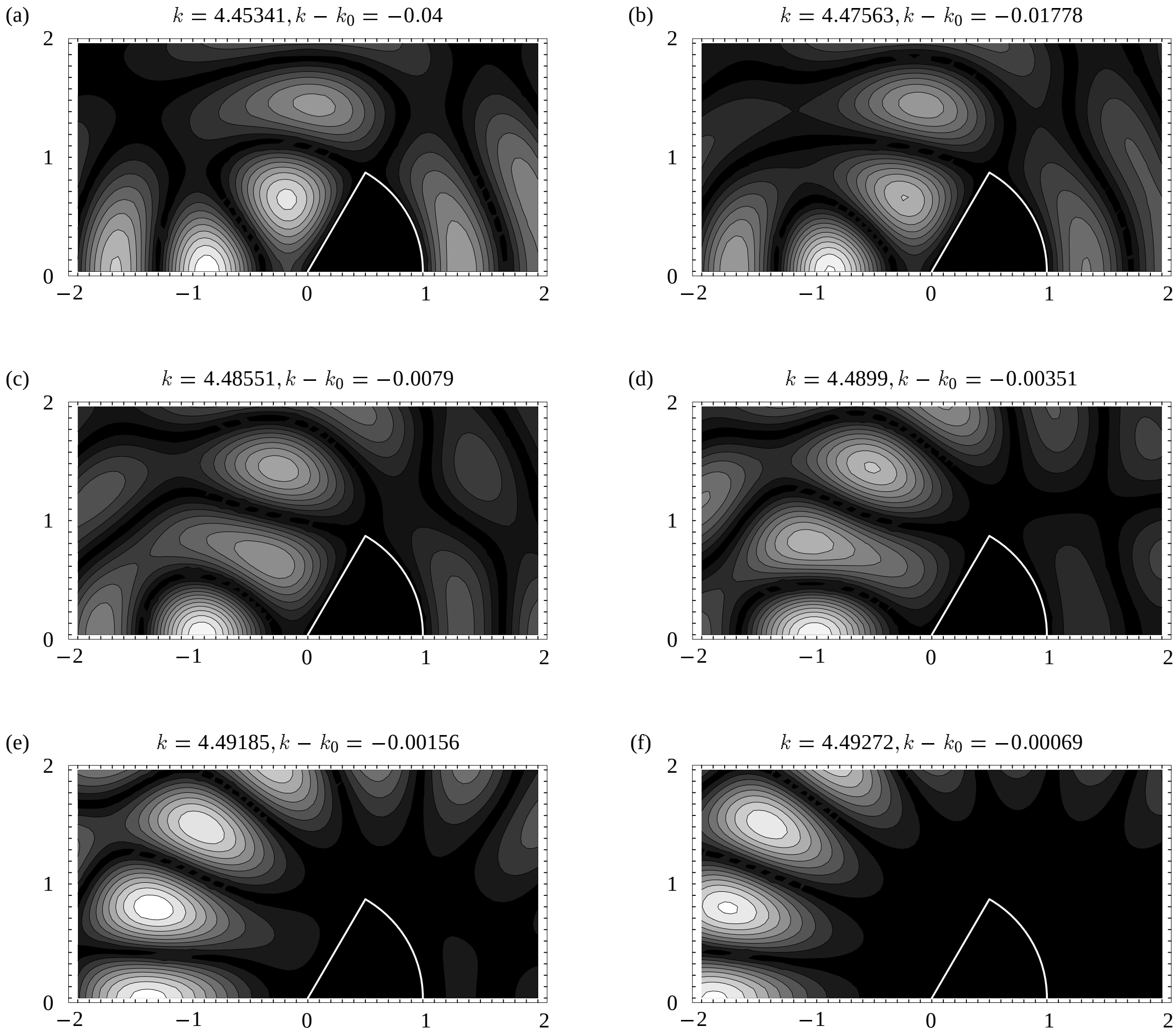

Fig. 8: Absolute value of the scattering solution corresponding to the phase which approaches $2\pi$ as $k \uparrow k_0$. Black corresponds to amplitudes below 0.05 and white to amplitudes above 1.95, with equal spacing between contours.

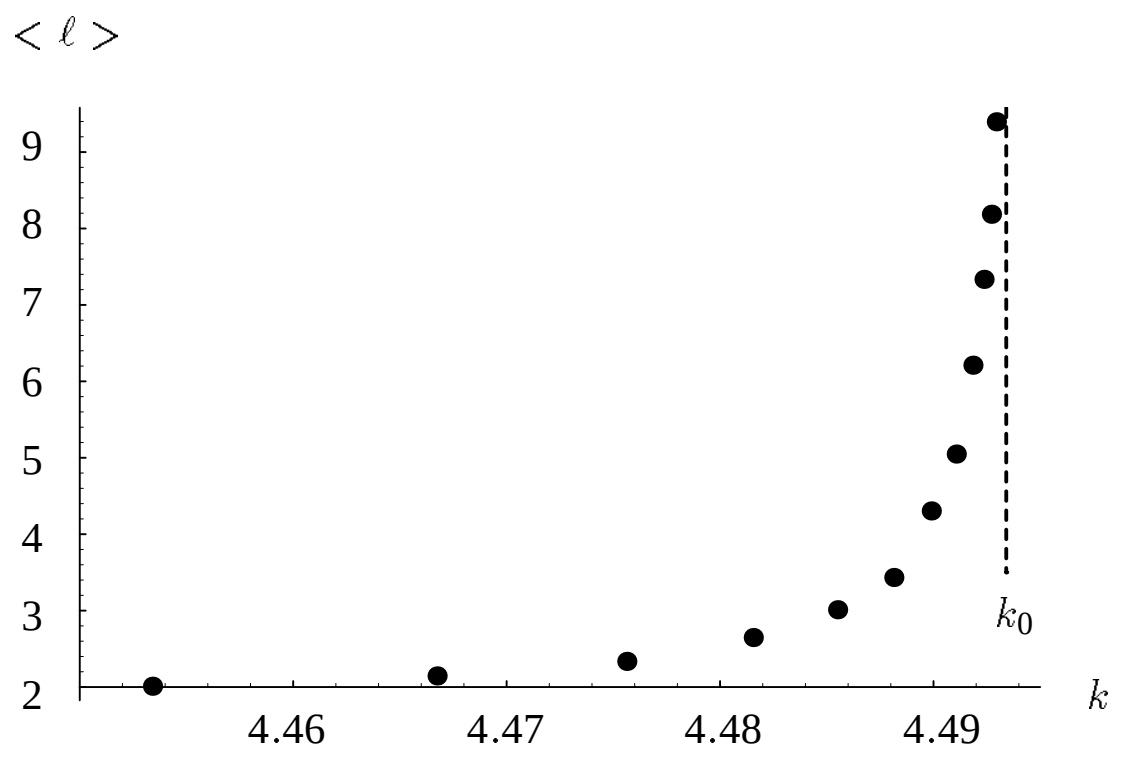

Fig. 9: Average angular momentum of the scattering solution as a function of $k \uparrow k_0$, as defined in Eq.(18).

angular momentum as a function of $k$. It is defined by

$$< \ell > = \sum_{\ell} \ell |v_\ell|^2 \,, \tag{18}$$

with $v$ as in the previous paragraph.

Second, in Fig. 8 we also see that the values of the eigenfunction near the obstacle go to 0 as $k \uparrow k_0$. This reflects the property that in this limit, the exterior eigenfunction goes (pointwise) to zero. The numerical studies indicate further that the semiclassical scattering phase $\theta_1(k)$ satisfies $\theta_1'(k) \downarrow 0$ as $k \uparrow k_0$. The derivative goes to zero faster than any power of $k_0 - k$.

As a final comment, we note that the phenomena which occur for the cake are believed to be generic. Indeed, we used the null field method to calculate the $S$-matrix for scattering from a square billiard and from a Stadium boundary. In both cases, the various features which were discussed above for the cake appear also for the other shapes. The results for the square billiard have been reported elsewhere [4]. The first 71 eigenenergies of the Stadium billiard (with a ratio of 1.8 between the length of the straight segment and the diameter of the circle) were calculated, using a truncated $S$-matrix of dimension $\Lambda = \Lambda_{\mathrm{sc}} + \Lambda_{\mathrm{e}}$, where $\Lambda_{\mathrm{sc}}$ is given by Eq.(8) and $\Lambda_{\mathrm{e}} \leq 7$. For $\Lambda_{\mathrm{e}} = 7$ their mean relative deviation from eigenvalues obtained using other numerical methods is $3 \cdot 10^{-5}$. Increasing $\Lambda_{\mathrm{e}}$ by 1 did not affect the first 6 significant digits of the calculated eigenenergies.

## 4 Discussion

The inside-outside duality which was discussed and illustrated in the present paper becomes particularly simple and intuitive when it is considered in the semiclassical domain. Let us consider first the *classical* description of the scattering process, which is entirely equivalent to geometrical optics. Consider a ray which hits the boundary from the outside. It is completely specified in terms of its direction $\phi_i$ and the angular momentum $\ell_i = \mathbf{k} \wedge \mathbf{r} = b_i k$, where $b_i$ is the impact parameter. The outgoing ray is reflected at an angle $\phi_f$ and with an angular momentum $\ell_f$. The physical scattering event is a “one shot” event. To turn this into a map we have to remember that $(\phi, \ell)$ defines actually a line, which has two branches, one incoming and the other outgoing. A scattering map can be defined by re-injecting the outgoing particle towards the scatterer along the incoming branch. This procedure can be iterated indefinitely [9, 11] thus generating an area preserving Poincaré scattering map. The domain of this map consists of all the values $(\phi, \ell)$ which correspond to lines which intersect the boundary. This is an annulus (of area $A = 2k|\Gamma|$ for a convex domain with boundary $\Gamma$) on the cylinder $0 \leq \phi < 2\pi$, $-\infty < \ell < \infty$. The complement of the annulus corresponds to $(\phi, \ell)$ values which define lines which do not intersect the boundary. Thus, the classical phase space is divided into two invariant domains: The relevant annulus where classical scattering occurs, and the remainder where the map is the identity $(\phi_f, \ell_f) = (\phi_i, \ell_i)$.

The $S$-matrix is the quantum (wave) analogue of the classical scattering map. In the semiclassical approximation, the structure of the $S$-matrix follows the strict partition of the classical phase space. The restriction of the $S$-matrix to the space of relevant angular momenta corresponds to the nontrivial classical scattering. In the complementary subspace, the $S$-matrix is the identity operator. The entries $S_{\ell,\ell'}$ where either $\ell$ or $\ell'$ are not relevant angular momenta correspond to transitions which are classically *forbidden*. The main difference between the semiclassical $S$-matrix and the exact one is that in the latter, tunneling and diffraction effects contribute to classically forbidden entries. However, these contributions are expected to be exponentially small in $1/k$, which is the typical behavior of tunneling contributions. As long as one is interested in the semiclassical domain, all that matters is the restriction of the $S$-matrix to the relevant subspace. This is a finite dimensional space and the quantization condition can be cast in the form det$(S(k) - I) = 0$. In this domain the inside-outside duality is observed in the strong sense. The intricate avoided crossings which necessitate the more delicate theory are due to genuine quantum mechanical effects. The smallness of these effects is the reason why one obtains reasonably precise values of the inside energies when one restricts the $S$-matrix to the "relevant" subspace, and why they are very accurate when the space is enlarged to include a few evanescent modes.

**Acknowledgments**.This work was supported by the Fonds National Suisse, the US-Israel Binational Science Foundation, and the Minerva Center for Nonlinear Physics of Complex Systems. BD acknowledges support from the Alexander-von-Humboldt foundation.